# Powering The Intra-cluster Filaments in Cool-Core Clusters of Galaxies

Gary J. Ferland[a]

[a]Physics, University of Kentucky, Lexington, KY 40506, USA

**Abstract.** The first radio surveys of the sky discovered that some large clusters of galaxies contained powerful sources of synchrotron emission. Optical images showed that long linear filaments with bizarre emission-line spectra permeated the intra-cluster medium. Recent observations in the infrared and radio show that these filaments have very strong emission lines of molecular hydrogen and carbon monoxide. The mass of molecular material is quite large, the gas is quite warm, and the filaments have not formed stars despite their ~Gyr age. I will discuss the general astrophysical context of large clusters of galaxies and how large masses of molecular gas can be heated to produce what we observe.

The unique properties of the filaments are a result of the unique environment. Magnetically confined molecular filaments are surrounded by the hot intra-cluster medium. Thermal particles with keV energies enter atomic and molecular regions and produce a shower of secondary non-thermal electrons. These secondaries collisionally heat, excite, dissociate, and ionize the cool gas. While ionization is dominated by these secondary particles, recombination is controlled by charge exchange, which produces the unusual optical emission line spectrum. I will describe some of the physical processes that are unique to this environment and outline some of the atomic physics issues.

**Keywords:** Galaxy clusters; atomic physics.
**PACS:** 98.58.Ay.

## INTRODUCTION

Despite the fact that the intracluster medium of galaxy clusters constitutes the largest reservoir of baryons in the universe, we do not understand how it sustains its high temperatures. A cooling flow, in which gas cools and condenses towards the center, was long expected. The *XMM Newton* and *Chandra* discovery that the gas does not cool efficiently below $5\times10^7$ K [1] suggests that a feedback process heats the intracluster medium. Interactions between the hot gas and radio lobes ejected from the central Active Galactic Nucleus (AGN; [2]) are one source of heating, although the details remain uncertain.

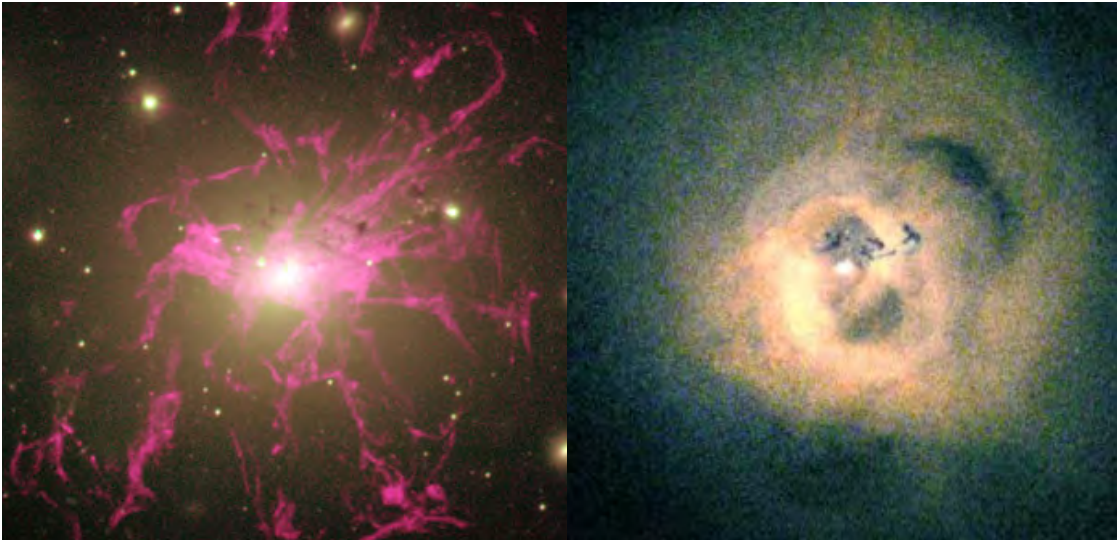

FIGURE 1. Optical (left) and X-ray (right) images centered on NGC 1275, the AGN in the Perseus cluster. The 5 keV plasma surrounds the atomic/molecular filaments.

Curiously, the central galaxies in "cool core clusters", where the AGN feedback process is most likely to operate, are frequently surrounded by networks of fine filaments [3]. NGC 1275, the central AGN in the Perseus cluster, is the best studied (Figure 1). The optical spectrum, dominated by low-ionization emission lines such as [O I], [O II], [N I] and [N II], is unlike any produced by H II regions or planetary nebulae. Infrared spectroscopy shows remarkably strong $H_2$ emission [4] while CO observations reveal that the filaments have a molecular mass as high as $4\times10^{10}$ solar masses [5]. This large reservoir of molecular gas shows that not all cooling is suppressed and may provide the fuel for future star formation and AGN activity in the central galaxy.

The origin of the filaments, and their role in the AGN feedback process, are central questions. They may represent material ejected from the central AGN or cooled from the surrounding hot plasma. Their composition would be one way to discriminate. If ejected from the AGN the abundances would be above solar while material cooled from the surrounding gas would share its subsolar composition.

Figure 2 shows the "horseshoe", a filament with extensive spectroscopy. X-ray [6], optical [7]; Spitzer [4], Herschel [8], and radio [9] observations reveal a wide range of ionization, from O VII in the X-ray, to low-ionization optical lines ([O II], [O I], [N II], and, curiously, the enigmatic [N I] λ5199 doublet), $H_2$ in the IR, and CO and HCN in the radio. The $H_2$ spectrum has a correlation between excitation potential and temperature [10], showing that the molecule exists in regions with a range of temperatures, extending up to the $H_2$ dissociation limit.

The right panel of Figure 2 shows the small-scale structure revealed in HST images [11]. Filaments have widths as small as 70 pc but lengths of tens of kpc. They are composed of remarkably thin and long clouds. The images are reminiscent of magnetic structures such as coronal loops. Strong magnetic fields, $B > 100$ mG, are needed to sustain this linear geometry [11] and could explain why such a large mass (approaching $10^{11}$ $M_{sun}$ [5]) can resist star formation for 1-5 Gyr [12].

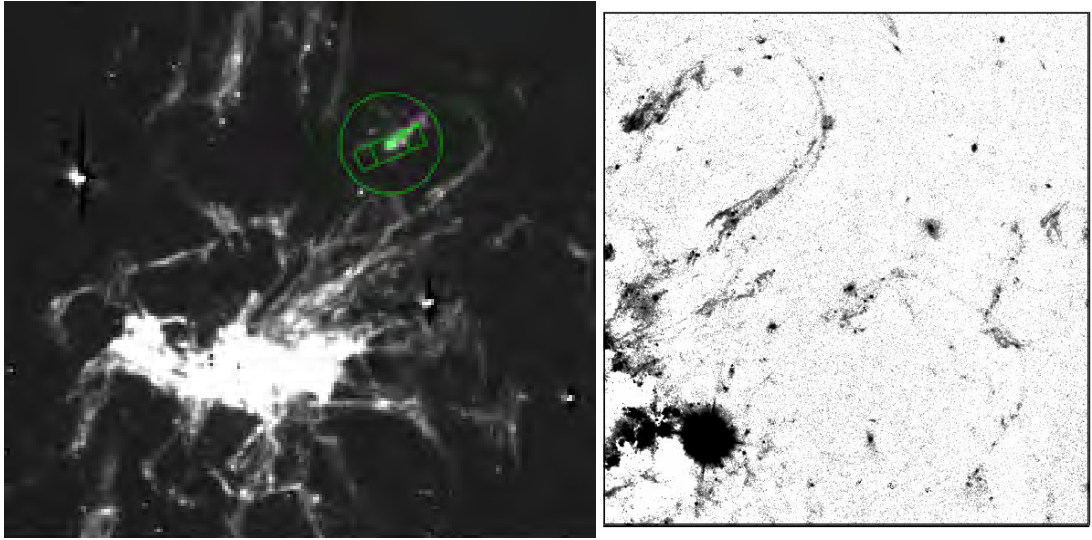

FIGURE 2 The Horseshoe filament in the Perseus cluster. The Hα image on the left shows NGC 1275 and the parts of the Horseshoe where the most extensive spectroscopy exists. The HST image on the right shows the Horseshoe in the upper left quadrant and reveals the fine structure that is present [11].

## THREE POSSIBLE HEATING SOURCES

The energy source for the filaments remains elusive and has been the focus of my recent work. The star-formation rates in the central galaxy are high but there is little evidence of star formation within the extended filaments. They exhibit a rich optical, IR, and radio emission line spectrum, and X-ray line emission is inferred [6]. The species present include highly ionized He-like oxygen, first and second ions of the abundant elements, and $H_2$, CO, and HNC [9]. It has long been known that O-star photoionization cannot produce the observed emission [13]. Some ways to produce strong $H_2$ emission are discussed in [14].

In general there are three possible sources of heat and ionization for interstellar clouds. These are discussed in turn in the following, with more details given in [16].

First, and the most important in galactic nebulae, is photoionization. Here a star, star cluster, accreting black hole, or a collapsed object such as a neutron star or white dwarf, produces energetic photons. Distant clouds are both heated and ionized by the resulting photoionization. The graduate text by Osterbrock and Ferland ([15]; hereafter AGN3) provides many of the details. Depending on the intensity of the ionizing radiation field, the gas may have a wide range of ionization and temperature, anywhere from cold and neutral up to hot with the heavy elements fully ionized.

Shocks are a second possibility. This happens, for instance, when an outflowing wind from a newly formed star encounters surrounding stationary material. A range of heating and ionization is created, which depends on the shock velocity and the role of the ambient magnetic field in cushioning its effects.

Ionizing particles are a third possibility. Many source of high-energy particles are present in the cluster environment. Cosmic rays, relativistic particles associated with the synchrotron jets, and the X-ray emitting intracluster medium are all certainly present. In situ particles, produced by transient MHD phenomena such as magnetic reconnection, are another possibility. The filaments' strong resemblances to coronal loops, which are powered this way, suggest this.

## AND THEIR ATOMIC PHYSICS

A typical filament spectrum is remarkably different from those of galactic nebulae. The optical spectrum has strong low-ionization lines such as [O II], [N II], [S II], and [O I]. The $H_2$ lines in the near IR are remarkably strong relative to hydrogen recombination lines, as discussed in [14]. The three energy sources mentioned above produce different effects on atomic or molecular gas. The remarkable spectrum can reveal the nature of the energy source that produced it.

Figure 3 shows the ionization and recombination rates for atoms and first ions of six elements seen in the optical spectrum. The models and atomic data are summarized in [16]. Photoionization produces low-energy photoelectrons, and there will be a significant number of free electrons in the gas. The photoelectrons rapidly undergo elastic collisions with other free electrons and, as a result, quickly share their energy with them. The result is that free electrons have a kinetic temperature that is small compared with the ionization potentials of the ions that are present.

Valence shell photoionization cross sections are typically of order several Megabarns, so the photoionization rates do not vary greatly. The rate is lower for He due to its higher ionization potential, but not dramatically so. The ionization ratio $A^+/A^0$ is proportional to the ratio of ionization to recombination rates. There will be a relatively uniform distribution of first ion populations, except for $He^+$, which will be present but only weakly so. Warm gas is only present when ions are present so essentially no [N I] emission is produced. This does not reproduce the observations.

The second class of possible heating sources adds energy to the free electrons as heat. Shocks are an example. Like photoionization this produces a well-defined kinetic temperature. Rates are shown in Figure 4. Ionization is by thermal collisional impact ionization. These rates have Boltzmann factors so essentially no $He^+$, with its very high ionization potential, is produced when the temperature is low enough to produce other first ions. Charge exchange is very important because of the relatively large abundance of $H^0$. This causes the large enhancement in rates for O, and to some extent, N. This model produces no He I emission (from recombination of $He^+$) and is ruled out.

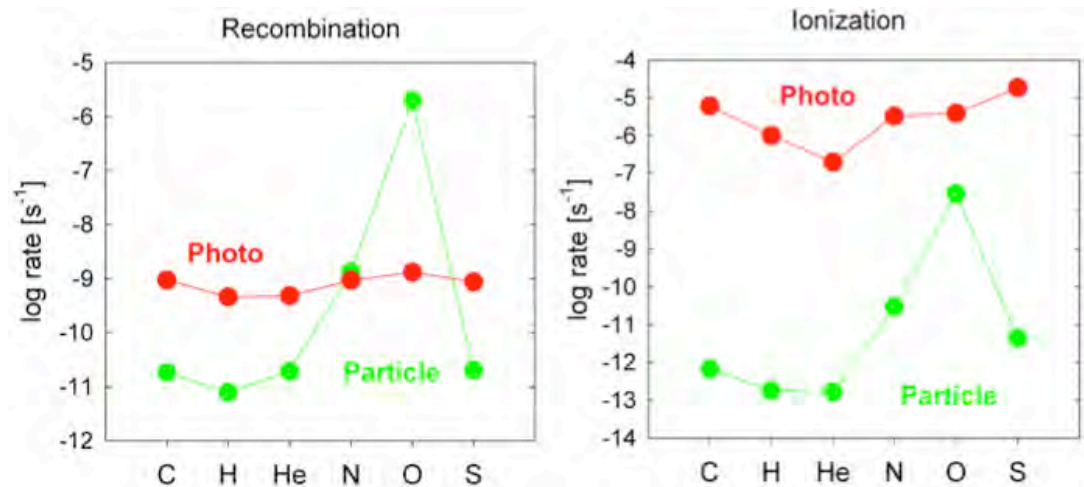

FIGURE 3. This shows the ionization and recombination rates for atoms / ions of six elements detected in the optical spectrum. An ionizing particle and photoionization model are compared. The ratio of the ion to atom densities is equal to the ratio of the ionization to recombination rates.

We have argued that ionizing particles produce the spectrum [16]. The essential physics is the following: higher energy ionizing particles enter atomic or molecular regions and lose their energy by a series of collisions that produce a large number of secondary suprathermal electrons. These secondaries simultaneously heat, ionize, and dissociate the gas. Most atoms have similar cross sections for high-energy collisional ionization so their ionization rates are comparable, as shown in Figures 3and4. The ions recombine by charge transfer, a quasi-molecular process that has vastly different rates for different ions [17]. The unusual mix of species is the result of secondary collisional ionization followed by charge-exchange recombination. He I and [N I] emission are both produced. Heating and dissociation by the suprathermal secondaries produces the $H_2$ spectrum. Ionizing particles account for the full optical/IR spectrum.

## IONIZING PARTICLES, MOLECULAR GAS, AND CONDUCTION FROM THE INTRACLUSTER MEDIUM

Our final ionizing-particle model is summarized in [16] and [18]. Magnetic field lines establish the geometry so that gas is free to move along these lines to maintain constant pressure. A variety of energy injection rates must be present to produce molecules, atoms, and ions. Gas moves along the field lines to maintain constant pressure. The model reproduces the full range of observations with a minimum of free parameters [16].

Whatever its source, a ~5 keV thermal particle entering cold molecular gas will produce the same knock-on secondary ionization effects as a cosmic ray, an example of how the same microphysics underlies macroscopically different regimes. [18]

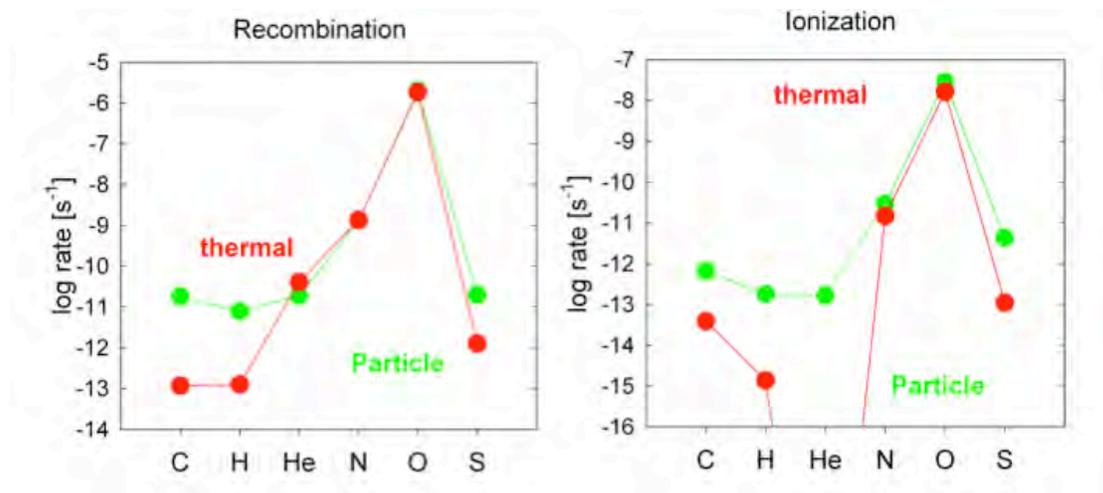

FIGURE 4. Similar to Figure 3, but a comparison of the thermal and particle heated models.

shows that the intracluster medium has sufficient energy flux to account for the energetics by thermal conduction. Ionizing particles penetrate the magnetically confined filaments by reconnection diffusion. The result is that the remarkable filament spectrum is the result of the remarkable conditions within cool-core clusters of galaxies with its magnetically confined molecular filaments surrounded by hot plasma.

ACKNOWLEDGEMENTS: My work on the microphysics of non-equilibrium gas is supported by the NSF (0908877; 1108928; and 1109061), NASA (07-ATFP07-0124, 10-ATP10-0053, and 10-ADAP10-0073), JPL (1430426), and STScI (HST-AR-12125.01 and HST-GO-12309).